\pdfoutput=1
\documentclass[man]{apa6}

\usepackage{amsmath}
\usepackage{pdfpages}
\usepackage[breaklinks]{hyperref}
\hypersetup{
  pdfborderstyle={/S/U/W 1},
  colorlinks = false,
  linkbordercolor = {1 1 1},
  urlbordercolor = {0 0.5 1},
  citebordercolor = {0 0.5 1}
} 
\usepackage[T1]{fontenc}
\usepackage{textcomp}
\usepackage[american]{babel}
\usepackage{txfonts}
\usepackage{setspace} 
\usepackage[natbibapa]{apacite}
\usepackage{url}
\usepackage{array}
\usepackage[low-sup]{subdepth}
\usepackage{booktabs}
\usepackage{dcolumn}
\usepackage{caption}
\usepackage{graphicx}
\usepackage{color}
\usepackage{inconsolata}

\newcommand\widebar[1]{%
  \hbox{%
    \vbox{%
      \hrule height 0.6pt%
      \kern0.256ex%
      \hbox{%
        \kern-0.15em%
        \ensuremath{#1}%
        \kern-0.1em%
      }%
    }%
  }%
} 

\DeclareMathSymbol{\beta}{\mathalpha}{lettersA}{12}
\DeclareMathSymbol{\delta}{\mathalpha}{lettersA}{14}
\DeclareMathSymbol{\epsilon}{\mathalpha}{lettersA}{15}
\DeclareMathSymbol{\varepsilon}{\mathalpha}{lettersA}{34}
\DeclareMathSymbol{\theta}{\mathalpha}{lettersA}{18}
\DeclareMathSymbol{\omega}{\mathalpha}{lettersA}{33}
\DeclareMathSymbol{\rho}{\mathalpha}{lettersA}{26}
\DeclareMathSymbol{\upsilon}{\mathalpha}{lettersA}{29}
\DeclareMathSymbol{\tau}{\mathalpha}{lettersA}{28}
\DeclareMathSymbol{\mu}{\mathalpha}{lettersA}{22}
\DeclareMathSymbol{\nu}{\mathalpha}{lettersA}{23}
\DeclareMathSymbol{\psi}{\mathalpha}{lettersA}{32}
\DeclareMathSymbol{\kappa}{\mathalpha}{lettersA}{20}
\DeclareMathSymbol{\lambda}{\mathalpha}{lettersA}{21}
\DeclareMathSymbol{\phi}{\mathalpha}{lettersA}{30}
\DeclareMathSymbol{\pi}{\mathalpha}{lettersA}{25}
\DeclareMathSymbol{\sigma}{\mathalpha}{lettersA}{27}
\DeclareMathSymbol{\alpha}{\mathalpha}{lettersA}{11}
\DeclareMathSymbol{\gamma}{\mathalpha}{lettersA}{13}
\DeclareMathSymbol{\chi}{\mathalpha}{lettersA}{31}
\DeclareMathSymbol{\eta}{\mathalpha}{lettersA}{17}
\DeclareMathSymbol{\zeta}{\mathalpha}{lettersA}{16}

\title{Multiple Imputation of Missing Covariate Values in Multilevel Models With Random Slopes: A Cautionary Note}
\shorttitle{Imputation of Covariates with Random Slopes}
\twoauthors{Simon Grund and Oliver L\"udtke}{Alexander Robitzsch}
\twoaffiliations{Leibniz Institute for Science and Mathematics Education;\\Centre for International Student Assessment}{Federal Institute for Education Research, Innovation and Development of the Austrian School System}

\authornote{\noindent This research was supported by a grant of the German Research Foundation (DFG), awarded to Oliver L\"udtke (LU 1636/1-1). Correspondence concerning this article should be addressed to Simon Grund, Leibniz Institute for Science and Mathematics Education, 24118 Kiel, Germany.

\noindent
Phone: +49 431 880 5653; E-mail: grund@ipn.uni-kiel.de}

\abstract{Multiple imputation (MI) has become one of the main procedures used to treat missing data, but guidelines from the methodological literature are not easily transferred to multilevel research. For models including random slopes, proper MI can be difficult, especially when covariate values are partially missing.
In the present paper, we discuss applications of MI in multilevel random coefficient models, theoretical challenges posed by slope variation, and current limitations of standard MI software.
Our findings from three simulation studies suggest that (a) MI is able to recover most parameters but is currently not well suited to capture slope variation entirely when covariate values are missing, (b) MI offers reasonable estimates for most parameters even in smaller samples or when its assumptions are not met, and (c) listwise can be an alternative worth considering when preserving the slope variance is particularly important.}


\begin{document}

\thispagestyle{empty}
This article is published in \emph{Behavior Research Methods}, and the print version is accessible using the link below. The copyright to the print version belongs to the journal. This manuscript is the author's personal manuscript after peer review, which was submitted to the journal upon acceptance. It is not the version of record and may not exactly replicate the print version.

~\\[1ex]

The print version of the article can be found online:

\url{http://link.springer.com/article/10.3758/s13428-015-0590-3}

~\\[1ex]

The correct citation for this article is:

Grund, S., L\"udtke, O., \& Robitzsch, A. (2016). Multiple imputation of missing covariate values in multilevel models with random slopes: A cautionary note. \emph{Behavior Research Methods}, \emph{48}, 640-649. doi: 10.3758/s13428-015-0590-3

\newpage
\setcounter{page}{1}

\maketitle
Multilevel data are often found in psychological research.
The complex pattern of variability in such data allows the use of statistical models that can accommodate multiple sources of variation. 
In recent years, multilevel models have become the standard tool for analyzing such data structures \citep{Raudenbush2002}.
In the social sciences, missing data (MD) represent a pervasive problem that has received considerable attention during the last two decades.
There is consensus in the methodological literature that methods such as multiple imputation (MI) are much better suited for treating missing data than traditional approaches such as listwise or pairwise deletion \citep{Little2002, Schafer2002}.

Although several book-length treatises have familiarized applied researchers with modern missing data methods \citep{Allison2001, Enders2010, Graham2012, vanBuuren2012}, less has been said about how to deal with missing values in multilevel research.
Previous studies concerned with missing data in multilevel modeling have consistently found that parameter estimates can be seriously distorted if the multilevel structure is not taken into account in the imputation process \citep{Andridge2011, vanBuuren2011a}.
However, these studies have focused on random intercept models, which assume that relations between variables do not vary across groups. 

In the present article, we focused on random slope models.
These models are frequently used in organizational and educational research to investigate whether relations at Level~1 (e.g., students, employees) vary across Level~2 units (e.g., classes, working teams) or in longitudinal research to assess different developmental trajectories across subjects.
For example, {\NoHyper{\citet{Hochweber2014}}} investigated the relationship between students' mathematics achievement and their math grades.
\citet{Hussong2008} examined the effects of parents' alcohol abuse on the development of children's internalizing behavior.

Using a multivariate mixed effects model and the software \emph{pan}, we explored strategies for dealing with missing data in models with random slopes.
In three simulation studies, we considered incomplete data on outcome variables, predictor variables, and both variables simultaneously, as well as different sample properties and patterns of missing data.

\subsection{Missing Data in Multilevel Research}
\subsubsection{The multivariate mixed effects model}
A few options for treating missing data in multilevel models are available in standard statistical software.
The pan package \citep{Schafer2014} has been recommended for MI of multilevel data \citep{Graham2012, Enders2010} and is easily accessible through the statistical software R \citep{RCoreTeam2014}.
The statistical model behind pan, which we will refer to as the \emph{pan model}, is the multivariate mixed effects model as presented by \citet{Schafer1997a}.
The pan model is capable of treating multilevel missing data but may also be used to describe both the imputation and analysis of multilevel data.
The model reads
\begin{equation} 
  \label{eq:pan}
  \mathbf{Y}_j = \mathbf{X}_j \boldsymbol\beta + \mathbf{Z}_j \mathbf{b}_j + \mathbf{E}_j \text{,}
\end{equation}
where $j = 1 , \ldots , G$ denotes groups or other observational units at Level~2.
Here, the response matrix $\mathbf{Y}_j$ of group $j$ is regressed on a design matrix $\mathbf{X}_j$ (containing intercept and predictor values) with associated \emph{fixed effects} $\boldsymbol\beta$ and a design matrix $\mathbf{Z}_j$ with associated group-specific \emph{random effects} $\mathbf{b}_j$.
The random effects matrix $\mathbf{b}_j$ (with columns stacked) is assumed to follow a normal distribution with mean zero and covariance matrix $\boldsymbol\Psi$ (iid. for all groups).
Each row of the error matrix $\mathbf{E}_j$ is assumed to follow a normal distribution with mean zero and covariance matrix $\boldsymbol\Sigma$ (iid. for all individuals).
Note that in the pan model, $\boldsymbol\Sigma$ is missing the index $j$ and is thus assumed to be the same for each group. 

Suppose our dataset consists of two variables $X$ and $Y$, both of which are Level~1 variables that have some variation at Level~2.
In a special application of the pan model, we may want to estimate the regression of $Y$ on $X$ with varying coefficients across groups, that is, the random coefficient model (RC model).
This model results if we write the outcome $Y$ (e.g., students' math grades) on the left-hand side of Equation \ref{eq:pan}, and the covariate $X$ (e.g., individual achievement) on the right-hand side, and allow for the intercepts and slopes to vary across groups.
We will also call the RC model the \emph{analyst's model} as it fits our supposed research question.
Finally, we can express the parameters of the analyst's model in a single expression $\boldsymbol\theta = (\boldsymbol\beta,\boldsymbol\Psi,\boldsymbol\Sigma)$ and write $f(Y|X,\boldsymbol\theta_X)$ in short for the RC model.

\subsubsection{Missing data terminology}
The common classification of missing data mechanisms found in \citet{Rubin1987} assumes a hypothetical complete data matrix, which is decomposed into observed and unobserved parts $\mathbf{Y}=(\mathbf{Y}_{obs},\mathbf{Y}_{mis})$ by an indicator matrix $\mathbf{R}$ denoting the missing data.
If values are missing as a random sample of the hypothetical complete data, that is, $P(\mathbf{R}|\mathbf{Y})=P(\mathbf{R})$, the data are missing completely at random (MCAR). If missingness depends on other variables but the data are MCAR with these partialled out, that is, $P(\mathbf{R}|\mathbf{Y})=P(\mathbf{R}|\mathbf{Y}_{obs})$, the data are missing at random (MAR).
These two missing data mechanisms are often called ``ignorable''. An ignorable missing data mechanism is highly beneficial for MI because all of the relevant information about the missing values is present in the dataset.
This is in contrast to data that are missing \emph{not} at random (MNAR) where missingness is additionally dependent on the missing part of the data, that is, $P(\mathbf{R}|\mathbf{Y})=P(\mathbf{R}|\mathbf{Y}_{obs}, \mathbf{Y}_{mis})$. For such ``nonignorable'' missingness, a general approach to an analysis of missing data is not feasible and strong assumptions have to be made about the missing data mechanism \citep{Carpenter2013}.

\subsubsection{Multiple imputation for multilevel models}
Multiple imputation, as introduced by \citet{Rubin1987}, is a convenient procedure for obtaining valid parameter estimates from partially unobserved data that usually relies on the MAR assumption (i.e., the observed values provide sufficient information about the missing data mechanism).
Using MI, the researcher draws independent random samples from the posterior predictive distribution of the missing values given the observed data and a statistical model, thus generating a number of complete datasets to use in further analyses.
The final parameter estimates can be obtained according to the rules described by \citet{Rubin1987} simply by averaging over the parameter estimates from all imputed datasets.
Applying MI can be subtle and need not always be the most practical choice in multilevel research \citep{Peters2012, Twisk2013} because its validity is subject to some further conditions.

First, with increasing variation and sample size at Level~2, it becomes necessary to include the multilevel structure in the imputation model.
Ignoring the multilevel structure using single-level MI may result in biased parameter estimates \citep{vanBuuren2011a, Taljaard2008}.
Second, the analyst's model has to be considered, and the imputation model must be specified accordingly \citep{Schafer2003, Meng1994}.
Broadly speaking, the imputation model must account for the complexity of the desired analysis.
If an imputation model is used that does not include variables or parameters relevant to the analyst (e.g., slope variance), then the analysis results will be biased.
And third, the imputation model must incorporate relevant information about the missing data process, that is, variables predictive of missing variables or of the missingness itself \citep{Carpenter2013}, to make the MAR assumption more plausible \citep{Collins2001}.
Satisfying these conditions can be cumbersome when varying slopes are of interest.
However, little is known about how the quality of parameter estimates in multilevel modeling is affected if one of these conditions is not met.

\subsubsection{Missing covariates in models with random slopes}
When only the outcome variable contains missing values, MI for a random coefficient model is straightforward.
The imputation model can be specified in pan by writing the outcome on the left-hand side of Equation \ref{eq:pan}, and the covariate with fixed and random effects on the right-hand side.
This is the previously mentioned RC model, denoted $f(Y|X,\boldsymbol\theta_X)$.
The imputation model is then equivalent to the analyst's model.

Fewer guidelines are available if a covariate contains missing values.
If the outcome is completely observed, then a \emph{reversed} imputation model may be used. For this model, the covariate is written on the left-hand side of Equation \ref{eq:pan} and the outcome on the right-hand side (with fixed and random effects).
We will refer to this as the reversed RC model and denote it $f(X|Y,\boldsymbol\theta_Y)$.
This model assumes slope variation but does so by regressing $X$ on $Y$, which might induce bias into the parameter estimation.
So far, pan has been recommended only for missing covariates whose effect is fixed across groups \citep{Schafer1997a}.
Alternatively, for a \emph{multivariate} imputation model, denoted $f(X,Y|\boldsymbol\theta_0)$, both variables could be written on the left-hand side of Equation \ref{eq:pan} with random intercepts for both variables.
Slope variation is ignored in this model, but in contrast to the conditional models (i.e., reversed and regular RC) it is able to account for multivariate patterns of missing data.
An additional description of these models can be found in Supplement A in the online supplemental materials. 
The supplemental online materials can be downloaded from \url{http://dx.doi.org/10.6084/m9.figshare.1206375} .

In three simulation studies, we assessed the performance of conditional and multivariate MI for random slope models.
Study 1, Study 2 and Study 3 examined cases in which missing values occurred on the outcome, the covariate, or both variables, respectively.
Study 1 attempted to replicate findings of previous research on partially observed outcome variables.
We expected both conditional MI and LD to provide approximately unbiased estimates if the outcome was MAR \citep{Little2002, Carpenter2013}.
Study 2 focused on missing covariate data.
We expected that the reversed model would recover most parameters of the RC model but that it might perform poorly for the slope variance.
Listwise deletion was expected to provide biased estimates with MAR and MNAR data.
Study 3 examined multivariate missing data.
We expected that multivariate MI would underestimate the slope variance but would recover most other parameters.
We expected the results for LD to be similar to the results from the second study.

\section{Study 1}
The first study compared the performance of LD, conditional MI, and multivariate MI when the only outcome had missing values.
For conditional MI, both the analyst's model $f(Y|X,\boldsymbol\theta_X)$ and the imputation model $g(Y|X,\boldsymbol\omega_X)$ were RC models where $\boldsymbol\omega_X$ took on the same role as $\boldsymbol\theta_X$ but denoted a distinct set of model parameters.
These models were equally complex and fit the clustered structure of the data.
Multivariate MI was set up as described earlier, and LD was applied by restricting the analysis to complete cases only.

\subsection{Simulation and Methods}
\subsubsection{Data generation and imposition of missing values}
Two standardized normal variables $X$ and $Y$ were simulated. Both varied at two levels as indicated by their intraclass correlations (ICCs) $\rho_X$ and $\rho_Y$, respectively.
The covariate $X$ was simulated from its within- and between-group portions $X^W \sim N(0,1-\rho_X)$ and $X^B \sim N(0,\rho_X)$, respectively. Then $Y$ was simulated conditionally on $X$ according to Equation \ref{eq:pan} with fixed effects $\boldsymbol\beta = (\beta_0, \beta_1)$, where $\beta_0$ was zero due to standardization.
The covariance matrix of random effects was $\boldsymbol\Psi = \left( \begin{smallmatrix} \psi_{11}^2 & 0 \\ 0 & \psi_{22}^2 \end{smallmatrix} \right)$. Thus, the intercepts and slopes were uncorrelated. The Level~1 residual variance was $\boldsymbol\Sigma = \sigma^2$.
The variables in this study were parametrized by their ICC rather than their actual variance components. Given the ICC and a slope variance $\psi_{22}^2$, the other variance components followed \citep[see][]{Snijders2012} as \\[-2.5ex]
\begin{equation}
  \label{eq:varcomp}
  \begin{aligned}
    \sigma^2 &= (1-\rho_Y) - \beta_1^2(1-\rho_X) - \psi_{22}^2(1-\rho_X) \\
    \psi_{11}^2 &= \rho_Y - \beta_1^2\rho_X - \psi_{22}^2\rho_X \text{.}
  \end{aligned}
\end{equation}
~\\[-2ex]\noindent Missing values on $Y$ were imposed using a linear model for the latent response variable $R^{*}$. Values in $Y$ were set to be missing if their respective $R^{*}>0$ according to \\[-2.5ex]
\begin{equation}
  \label{eq:mismech}
  \begin{aligned}
    R^{*} &= \alpha + \lambda_1 X + \lambda_2 Y + \varepsilon_{R^{*}} \text{,}
  \end{aligned}
\end{equation}
where $\alpha$ is a value of the standard normal distribution according to a missing data probability (e.g., $\alpha = -0.67$ for 25\% missing data), and $\lambda_1$ and $\lambda_2$ are used to control the missing data mechanism. The residuals were distributed normally with mean zero and variance \\[-2ex]
\begin{equation}
  \label{eq:resmis}
    \sigma_{R^{*}}^2 = 1 - \lambda_1^2 - \lambda_2^2 - 2\;\! \lambda_1 \lambda_2 \text{Cov}(X,Y) \text{.}
\end{equation}
Table \ref{tab:study} provides an overview of the conditions included in all three studies. 
The two ICCs were set to be equal, that is, $\rho_X=\rho_Y=\rho$.
In order for $Y$ to be MCAR, we set $\lambda_1=\lambda_2=0$, and for MAR, we set $\lambda_1=0.5$ and $\lambda_2=0$.
For $Y$ to be MNAR, we chose equal values for $\lambda_1$ and $\lambda_2$ such that the error variance in $R^{*}$ was the same as in the MAR condition.
Hence, with $\text{Cov}(X,Y)=\beta_1=0.5$, we had $\lambda_1=\lambda_2=\sqrt{0.25/3} \approx 0.289$. 
The conditions were chosen to mimic typical data in psychology and the behavioral sciences \citep{Aguinis2013, Mathieu2012, Murray2003}.
\begin{table}[tbp]
  \begin{threeparttable}
    \renewcommand{\arraystretch}{1.35}
    \caption{Simulation Designs of Study 1, Study 2, and Study 3}
    \label{tab:study}
    \small
    \begin{tabular}{llll} \hline \\[-3.5ex]
    Design conditions& Study 1            & Study 2            & Study 3 \\[0.5ex] \hline
    Number of groups & 50, 150            & 50, 150            & 50, 150 \\
    Group size       & 10, 30             & 10, 30             & 10, 30 \\
    ICC              & .05, .15, .25      & .05, .15, .25      & .05, .15, .25 \\
    Fixed slope      & .50                & .50                & .50 \\
    Slope variance   & .01, .05, .10, .20 & .01, .05, .10, .20 & .01, .05, .10, .20 \\
    MD pattern       & univariate $Y$     & univariate $X$     & multivariate $X$ and $Y$ \\
    MD proportion    & 25\%, 50\%         & 25\%, 50\%         & 25\%, 50\% \\
    MD mechanism     & MCAR, MAR, MNAR    & MCAR, MAR, MNAR    & MCAR, MAR, MNAR \\
    Imputation models& RC model $f(Y|X,\boldsymbol\theta_X)$, 
                     & reversed RC $f(X|Y,\boldsymbol\theta_Y)$,
                     & multivariate $f(X,Y|\boldsymbol\theta_0)$ \\
                     & multivariate $f(X,Y|\boldsymbol\theta_0)$
                     & multivariate $f(X,Y|\boldsymbol\theta_0)$ & \\ \hline
    \end{tabular}
    \begin{tablenotes}[para,flushleft] ~\\[-1.5ex]
      {\small \textit{Note.} ICC = intraclass correlation of $X$ and $Y$; MD = missing data; MCAR = missing completely at random; MAR = missing at random; MNAR = missing not at random; RC = random coefficients.}
    \end{tablenotes}
  \end{threeparttable}
\end{table}

In summary, each simulated setting was defined by the number of groups (G), the number of individuals within each group (N), the ICC of $X$ and $Y$ ($\rho$), the fixed slope ($\beta_1$), the slope variance ($\psi_{22}^2$), the proportion of missing data, and the missing data mechanism (including the missing data effects $\lambda_1$ and $\lambda_2$).
Each setting was replicated 1,000 times.

\subsubsection{Imputation and data analysis}
The R package pan was used to impute missing values \citep{Schafer2014}.
We let pan perform 10,000 burn-in cycles before drawing one imputed dataset for every 200 cycles, leading to $M=50$ imputed datasets and 20,000 cycles in total \citep[see][]{Graham2007}.
Diagnostic plots regarding the convergence behavior of pan's Gibbs sampler are presented in Supplement B in the online supplemental materials.

Least-informative inverse-Wishart priors for $\boldsymbol\Sigma$ and $\boldsymbol\Psi$ were chosen with $\boldsymbol\Sigma \sim W^{-1}(\mathbf{I}_1,1)$ and $\boldsymbol\Psi \sim W^{-1}(\mathbf{I}_2,2)$ for conditional MI, and $\boldsymbol\Sigma \sim W^{-1}(\mathbf{I}_2,2)$ and $\boldsymbol\Psi \sim W^{-1}(\mathbf{I}_2,2)$ for multivariate MI, where $\mathbf{I}_n$ denotes the identity matrix of size $n$.
We fit the analyst's model to each imputed dataset using the R package lme4 \citep{Bates2013}.
The final parameter estimates were obtained according to Rubin's (1987) rules.
We note that choosing least-informative priors implies a prior expectation of variances of $.50$, which might induce bias into small variance components.
However, because non-informative priors are often desirable for MI, the same priors were used throughout the three studies.
Possible alternative specifications of the prior distribution will be reviewed in the General Discussion.
The computer code for running conditional and multivariate MI, with least-informative or alternative priors, is provided in Supplement C of the supplemental online materials.

Bias and the root-mean-square error (RMSE) were calculated for each condition and each parameter.
The bias is the mean difference between a parameter estimate $\hat \theta$ and its true value $\theta$, and is crucial for statistical reasoning in general.
The RMSE is the root of the mean squared difference between $\hat \theta$ and $\theta$ and represents both accuracy and precision (i.e., the variability) of an estimator.
Thus, it is an important measure of practical utility.

\subsection{Results and Discussion}
\begin{table}[tbp]
  \begin{threeparttable}
  \caption{Study 1: Bias and RMSE for Estimates Obtained from LD and MI Given Small Variance Components, Smaller or Larger Samples, and Missing $Y$ ~\\[2.0ex]}
    \label{tab:res1}
    \setlength{\tabcolsep}{5pt}
    \renewcommand{\arraystretch}{1.15}
    \small
    \begin{tabular}{lcccccccccccc}
\\[-5ex]\hline\\[-2.5ex]
 & \multicolumn{6}{c}{Bias} & \multicolumn{6}{c}{RMSE} \\ \cmidrule(lr){2-7}\cmidrule(lr){8-13}
 & \multicolumn{3}{c}{MCAR} & \multicolumn{3}{c}{MAR} & \multicolumn{3}{c}{MCAR} & \multicolumn{3}{c}{MAR} \\ \cmidrule(lr){2-4}\cmidrule(lr){5-7}\cmidrule(lr){8-10}\cmidrule(lr){11-13}
Est.  & LD & MV & RC & LD & MV & RC & LD & MV & RC & LD & MV & \multicolumn{1}{c}{RC} \\ 
[0ex]\hline\\[-2.8ex]
& \multicolumn{11}{c}{A: N=10, G=50, ICC=0.05, SV=0.01} \\[0ex] \hline \\[-2.7ex]$\beta_0^{}$  & $\phantom{-}.002$ & $\phantom{-}.002$ & $\phantom{-}.002$ & $\phantom{-}.003$ & $\phantom{-}.001$ & $\phantom{-}.003$ & $\phantom{-}.003$ & $\phantom{-}.003$ & $\phantom{-}.003$ & $\phantom{-}.003$ & $\phantom{-}.003$ & $\phantom{-}.003$ \\
$\beta_1^{}$  & $-.001$ & $-.002$ & $-.001$ & $\phantom{-}.002$ & $-.001$ & $\phantom{-}.002$ & $\phantom{-}.002$ & $\phantom{-}.002$ & $\phantom{-}.002$ & $\phantom{-}.002$ & $\phantom{-}.002$ & $\phantom{-}.002$ \\
$\psi_{11}^2$  & $-.005$ & $\mathbf{\phantom{-}.014}$ & $\phantom{-}.010$ & $-.002$ & $\mathbf{\phantom{-}.013}$ & $\phantom{-}.005$ & $\phantom{-}.001$ & $\phantom{-}.001$ & $\phantom{-}.001$ & $\phantom{-}.001$ & $\phantom{-}.001$ & $\phantom{-}.001$ \\
$\psi_{22}^2$  & $\mathbf{\phantom{-}.006}$ & $\phantom{-}.001$ & $\mathbf{\phantom{-}.023}$ & $\mathbf{\phantom{-}.007}$ & $\phantom{-}.001$ & $\mathbf{\phantom{-}.032}$ & $\phantom{-}.000$ & $\phantom{-}.000$ & $\mathbf{\phantom{-}.001}$ & $\phantom{-}.000$ & $\phantom{-}.000$ & $\mathbf{\phantom{-}.001}$ \\
$\psi_{12}^{}$  & $-.001$ & $-.002$ & $-.003$ & $\phantom{-}.000$ & $\phantom{-}.001$ & $\phantom{-}.008$ & $\phantom{-}.000$ & $\phantom{-}.000$ & $\phantom{-}.000$ & $\phantom{-}.000$ & $\phantom{-}.000$ & $\mathbf{\phantom{-}.000}$ \\
$\sigma_{}^2$  & $-.001$ & $-.004$ & $-.009$ & $-.007$ & $-.006$ & $-.008$ & $\phantom{-}.003$ & $\phantom{-}.003$ & $\phantom{-}.003$ & $\phantom{-}.003$ & $\phantom{-}.003$ & $\phantom{-}.004$ \\
[0ex]\hline\\[-2.8ex] 
& \multicolumn{11}{c}{B: N=30, G=150, ICC=0.05, SV=0.01} \\[0ex] \hline \\[-2.7ex]$\beta_0^{}$  & $-.000$ & $-.000$ & $-.000$ & $\phantom{-}.001$ & $\phantom{-}.000$ & $\phantom{-}.000$ & $\phantom{-}.000$ & $\phantom{-}.000$ & $\phantom{-}.000$ & $\phantom{-}.000$ & $\phantom{-}.000$ & $\phantom{-}.000$ \\
$\beta_1^{}$  & $\phantom{-}.000$ & $\phantom{-}.000$ & $\phantom{-}.000$ & $-.001$ & $-.001$ & $-.001$ & $\phantom{-}.000$ & $\phantom{-}.000$ & $\phantom{-}.000$ & $\phantom{-}.000$ & $\phantom{-}.000$ & $\phantom{-}.000$ \\
$\psi_{11}^2$  & $-.002$ & $\phantom{-}.001$ & $\phantom{-}.002$ & $-.000$ & $\phantom{-}.002$ & $\phantom{-}.001$ & $\phantom{-}.000$ & $\phantom{-}.000$ & $\phantom{-}.000$ & $\phantom{-}.000$ & $\phantom{-}.000$ & $\phantom{-}.000$ \\
$\psi_{22}^2$  & $\phantom{-}.000$ & $\mathbf{-.004}$ & $\mathbf{\phantom{-}.006}$ & $\phantom{-}.000$ & $\mathbf{-.005}$ & $\mathbf{\phantom{-}.009}$ & $\phantom{-}.000$ & $\phantom{-}.000$ & $\mathbf{\phantom{-}.000}$ & $\phantom{-}.000$ & $\phantom{-}.000$ & $\mathbf{\phantom{-}.000}$ \\
$\psi_{12}^{}$  & $-.000$ & $-.000$ & $-.000$ & $\phantom{-}.000$ & $-.000$ & $\phantom{-}.003$ & $\phantom{-}.000$ & $\phantom{-}.000$ & $\phantom{-}.000$ & $\phantom{-}.000$ & $\phantom{-}.000$ & $\phantom{-}.000$ \\
$\sigma_{}^2$  & $\phantom{-}.002$ & $\phantom{-}.005$ & $-.002$ & $\phantom{-}.000$ & $\phantom{-}.004$ & $-.001$ & $\phantom{-}.000$ & $\phantom{-}.000$ & $\phantom{-}.000$ & $\phantom{-}.000$ & $\phantom{-}.000$ & $\phantom{-}.000$ \\
[0ex]\hline\\[-2.8ex] 
\end{tabular}

    \begin{tablenotes}[para,flushleft] ~\\[-2.0ex]
      {\small \textit{Note.} LD = listwise deletion; MV = multivariate imputation; RC = conditional imputation (random coefficients); MCAR = missing completely at random; MAR = missing at random; $\beta_0$ = fixed intercept; ; $\beta_1$ = fixed slope; $\psi_{11}^2$ = intercept variance; $\psi_{22}^2$ = slope variance; $\psi_{12}$ = intercept-slope covariance; $\sigma^2$ = Level~1 residual variance.} 
    \end{tablenotes}
  \end{threeparttable}
\end{table}

Due to the large simulation design, only the most important findings will be reported.
Furthermore, only results for 25\% missing data will be reported as higher rates did not yield interesting results.
The complete results for Study 1 are given in Supplement D in the online supplemental materials.
Table \ref{tab:res1} shows the results of the first study for samples that featured small variance components (i.e., $\text{ICC}=.05$, $\psi_{22}^2=.01$) for MCAR and MAR data in smaller (N~=~10, G~=~50) and larger samples (N~=~30, G~=~150).
Notable values for bias and RMSE are presented in bold.
Bias presented in bold is at least $\pm5\%$ off the true value for fixed effects, and $\pm30\%$ off for variance components.
For parameters whose true value was zero, a threshold of $\pm.05$ was used.
For each simulated condition, the highest RMSE is printed in bold as long as it was significantly larger than that found for the complete datasets (at least twice as large).

As can be seen in Table \ref{tab:res1}, neither LD nor MI produced strongly biased results, but bias emerged under specific conditions for both MI procedures.
The multivariate imputation model underestimated the slope variance by as much as 50\% unless it was essentially zero (i.e., $.01$), but overestimated the intercept variance.
Conditional MI (using the RC model) overestimated both the intercept and slope variance (Table \ref{tab:res1}, top panel).
A sufficient sample size reduced bias to acceptable proportions even for the smallest variance components (Table \ref{tab:res1}, bottom panel).
For larger values of the ICC (i.e., $.15$ and $.25$) and the slope variance (i.e., $.05$, $.10$, and $.20$), this bias was reduced to essentially zero (see Supplement D).
Using LD, the intercept and slope variance were sometimes biased when samples were not sufficiently large.
When data were MNAR, all approaches yielded biased results (see Supplement D).

Listwise deletion has previously been shown to provide essentially unbiased estimates when the outcome is ignorably missing \citep[e.g.,][]{Little2002}.
Surprisingly, the imputation models overestimated small random effects variances in small samples.
We argue that this is a side effect of the least-informative prior which expects variances to be larger, and that bias may be reduced to zero when the prior is set on an appropriate scale (see general discussion).
From the data at hand, both LD and conditional MI can be recommended for univariate missing data on $Y$ provided that the sample is sufficiently large or the prior is set on an appropriate scale.
Care should be taken when small variance components are to be estimated, as overly non-informative priors may inflate them.
The multivariate model is useful if the slope variance is close to zero.

\section{Study 2}
The second study examined the performance of MI and LD with missing values on the covariate $X$.
The analyst's model was again the RC model $f(Y|X,\boldsymbol\theta_X)$, whereas conditional MI was carried out using the \emph{reversed} RC model $g(X|Y,\boldsymbol\omega_Y)$.
The two models fit the clustered structure of the data but differed in the way the slope variability was attributed.
Multivariate MI and LD were administered as before.

\subsection{Simulation and Methods}
The same procedures as applied in Study 1 were used to simulate data and impose missing values on the covariate $X$, whereas MAR was now dependent on the outcome $Y$.
Imputations were created by pan using the least-informative priors as chosen in Study 1.
The analyst's model was fit using lme4, and the bias and RMSE were calculated for each parameter in each setting.

\subsection{Results} 

\begin{table}[tbp]
  \begin{threeparttable}
    \caption{Study 2: Bias and RMSE for Estimates Obtained from LD and MI Given Small Variance Components, Smaller or Larger Samples, and Missing $X$ ~\\[1.8ex]}
    \label{tab:res2}
    \setlength{\tabcolsep}{5pt}
    \renewcommand{\arraystretch}{1.15}
    \small
    \begin{tabular}{lcccccccccccc}
\\[-5ex]\hline\\[-2.5ex]
 & \multicolumn{6}{c}{Bias} & \multicolumn{6}{c}{RMSE} \\ \cmidrule(lr){2-7}\cmidrule(lr){8-13}
 & \multicolumn{3}{c}{MCAR} & \multicolumn{3}{c}{MAR} & \multicolumn{3}{c}{MCAR} & \multicolumn{3}{c}{MAR} \\ \cmidrule(lr){2-4}\cmidrule(lr){5-7}\cmidrule(lr){8-10}\cmidrule(lr){11-13}
Est.  & LD & MV & RC & LD & MV & RC & LD & MV & RC & LD & MV & \multicolumn{1}{c}{RC} \\ 
[0ex]\hline\\[-2.8ex]
& \multicolumn{11}{c}{A: N=10, G=50, ICC=0.05, SV=0.01} \\[0ex] \hline \\[-2.7ex]$\beta_0^{}$  & $-.001$ & $-.001$ & $-.001$ & $\mathbf{-.161}$ & $-.000$ & $-.003$ & $\phantom{-}.003$ & $\phantom{-}.002$ & $\phantom{-}.002$ & $\mathbf{\phantom{-}.028}$ & $\phantom{-}.002$ & $\phantom{-}.002$ \\
$\beta_1^{}$  & $\phantom{-}.002$ & $-.002$ & $-.010$ & $\mathbf{-.042}$ & $-.002$ & $-.014$ & $\phantom{-}.002$ & $\phantom{-}.002$ & $\phantom{-}.002$ & $\mathbf{\phantom{-}.004}$ & $\phantom{-}.002$ & $\phantom{-}.002$ \\
$\psi_{11}^2$  & $-.005$ & $\phantom{-}.000$ & $-.003$ & $-.008$ & $\phantom{-}.002$ & $-.000$ & $\phantom{-}.001$ & $\phantom{-}.001$ & $\phantom{-}.001$ & $\phantom{-}.001$ & $\phantom{-}.001$ & $\phantom{-}.001$ \\
$\psi_{22}^2$  & $\mathbf{\phantom{-}.007}$ & $\phantom{-}.002$ & $\mathbf{\phantom{-}.008}$ & $\mathbf{\phantom{-}.005}$ & $\phantom{-}.002$ & $\mathbf{\phantom{-}.010}$ & $\phantom{-}.000$ & $\phantom{-}.000$ & $\phantom{-}.000$ & $\phantom{-}.000$ & $\phantom{-}.000$ & $\phantom{-}.000$ \\
$\psi_{12}^{}$  & $-.001$ & $-.002$ & $-.002$ & $-.002$ & $-.003$ & $-.006$ & $\phantom{-}.000$ & $\phantom{-}.000$ & $\phantom{-}.000$ & $\phantom{-}.000$ & $\phantom{-}.000$ & $\phantom{-}.000$ \\
$\sigma_{}^2$  & $-.001$ & $\phantom{-}.000$ & $\phantom{-}.000$ & $-.063$ & $\phantom{-}.000$ & $-.003$ & $\phantom{-}.003$ & $\phantom{-}.003$ & $\phantom{-}.003$ & $\mathbf{\phantom{-}.007}$ & $\phantom{-}.003$ & $\phantom{-}.003$ \\
[0ex]\hline\\[-2.8ex] 
& \multicolumn{11}{c}{B: N=30, G=150, ICC=0.05, SV=0.01} \\[0ex] \hline \\[-2.7ex]$\beta_0^{}$  & $-.000$ & $-.000$ & $-.000$ & $\mathbf{-.160}$ & $-.001$ & $-.002$ & $\phantom{-}.001$ & $\phantom{-}.000$ & $\phantom{-}.000$ & $\mathbf{\phantom{-}.026}$ & $\phantom{-}.000$ & $\phantom{-}.000$ \\
$\beta_1^{}$  & $-.000$ & $-.000$ & $-.003$ & $\mathbf{-.045}$ & $-.001$ & $-.005$ & $\phantom{-}.000$ & $\phantom{-}.000$ & $\phantom{-}.000$ & $\mathbf{\phantom{-}.002}$ & $\phantom{-}.000$ & $\phantom{-}.000$ \\
$\psi_{11}^2$  & $-.001$ & $-.001$ & $-.001$ & $-.006$ & $\phantom{-}.000$ & $\phantom{-}.001$ & $\phantom{-}.000$ & $\phantom{-}.000$ & $\phantom{-}.000$ & $\phantom{-}.000$ & $\phantom{-}.000$ & $\phantom{-}.000$ \\
$\psi_{22}^2$  & $-.000$ & $-.003$ & $-.000$ & $-.001$ & $-.003$ & $\phantom{-}.000$ & $\phantom{-}.000$ & $\phantom{-}.000$ & $\phantom{-}.000$ & $\phantom{-}.000$ & $\phantom{-}.000$ & $\phantom{-}.000$ \\
$\psi_{12}^{}$  & $-.000$ & $-.000$ & $-.000$ & $-.001$ & $-.000$ & $-.001$ & $\phantom{-}.000$ & $\phantom{-}.000$ & $\phantom{-}.000$ & $\phantom{-}.000$ & $\phantom{-}.000$ & $\phantom{-}.000$ \\
$\sigma_{}^2$  & $\phantom{-}.002$ & $\phantom{-}.005$ & $\phantom{-}.003$ & $-.060$ & $\phantom{-}.004$ & $\phantom{-}.002$ & $\phantom{-}.000$ & $\phantom{-}.000$ & $\phantom{-}.000$ & $\mathbf{\phantom{-}.004}$ & $\phantom{-}.000$ & $\phantom{-}.000$ \\
[0ex]\hline\\[-2.8ex] 
\end{tabular}

    \begin{tablenotes}[para,flushleft] ~\\[-2.0ex]
      {\small \textit{Note.} LD = listwise deletion; MV = multivariate imputation; RC = conditional imputation (random coefficients); MCAR = missing completely at random; MAR = missing at random; $\beta_0$ = fixed intercept; ; $\beta_1$ = fixed slope; $\psi_{11}^2$ = intercept variance; $\psi_{22}^2$ = slope variance; $\psi_{12}$ = intercept-slope covariance; $\sigma^2$ = Level~1 residual variance.} 
    \end{tablenotes}
  \end{threeparttable}
\end{table}
The results of Study 2 are reported in full in Supplement D.
Here, we will report the most important findings.
Table \ref{tab:res2} provides a brief overview of the results for samples that featured small variance components.
Estimating the fixed effects of the RC model proved to be more accurate and efficient using MI.
Specific difficulties emerged again for small variance components, that is, when samples featured small ICCs or little slope variation.
In contrast to when data were missing on $Y$, however, estimates of larger slope variances were not necessarily unbiased.

\subsubsection{Fixed effects}
As shown in Table \ref{tab:res2}, LD led to biased estimates for the fixed effects unless the data were MCAR (see Supplement D).
Bias for the fixed intercept varied between $-.098$ and $-.161$ with MAR data and between $-.055$ and $-.101$ with MNAR data.
The fixed slope was underestimated by approximately 6-10\% when the data were not MCAR.
Results from MI were essentially unbiased, but the reversed model exhibited a small downward bias across conditions.
The RMSE suggested that estimates obtained from MI were at least as efficient as those obtained by LD across conditions, and more efficient when data were not MCAR.

\begin{figure}
  \centering
  \includegraphics[width=\linewidth]{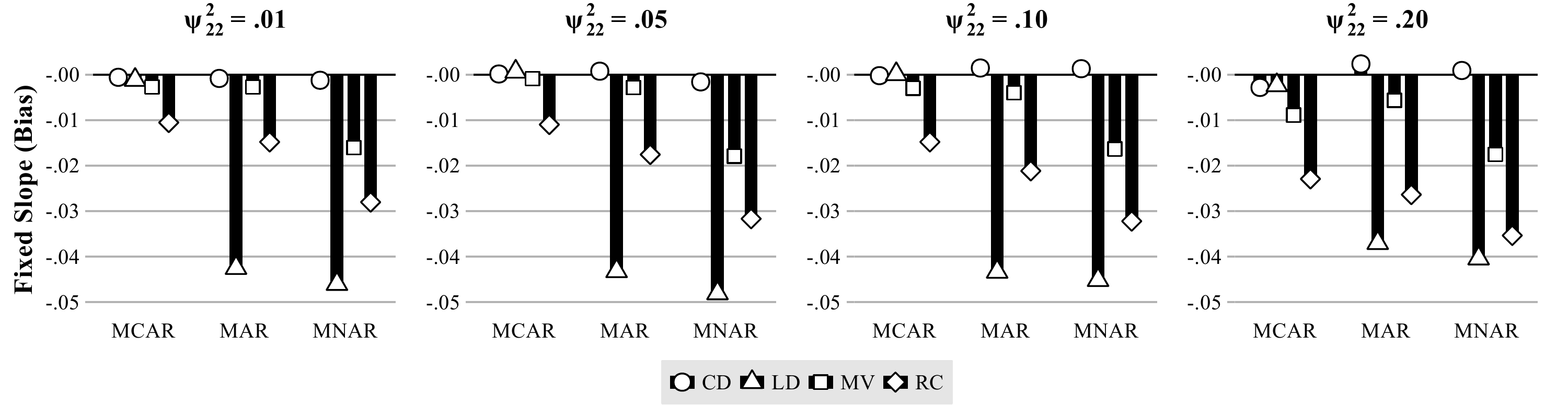}
  \caption{Bias in estimating the fixed slope for univariate missing data on $X$ (Study 2) for different MD mechanisms, MD techniques, and different amounts of slope variance. $\psi_{22}^2$ = true slope variance; MCAR = missing completely at random; MAR = missing at random; MNAR = missing not at random; CD = complete data; LD = listwise deletion; MV = multivariate imputation; RC = conditional imputation using the reversed RC model.}
  \label{fig:fixslo}
\end{figure}
Interestingly, bias from both LD and conditional MI was dependent on the amount of slope variation that was present in the dataset.
As slope variation increased, bias became weaker with LD, and stronger with conditional MI.
This result is illustrated in Figure \ref{fig:fixslo} for small samples (N~=~10, G~=~150), moderate ICCs (i.e., .15), and MAR data.
Nonetheless, estimates obtained from MI were more accurate and efficient across all conditions.

\subsubsection{Variance and covariance of random effects}
\begin{figure}
  \centering
  \includegraphics[width=\linewidth]{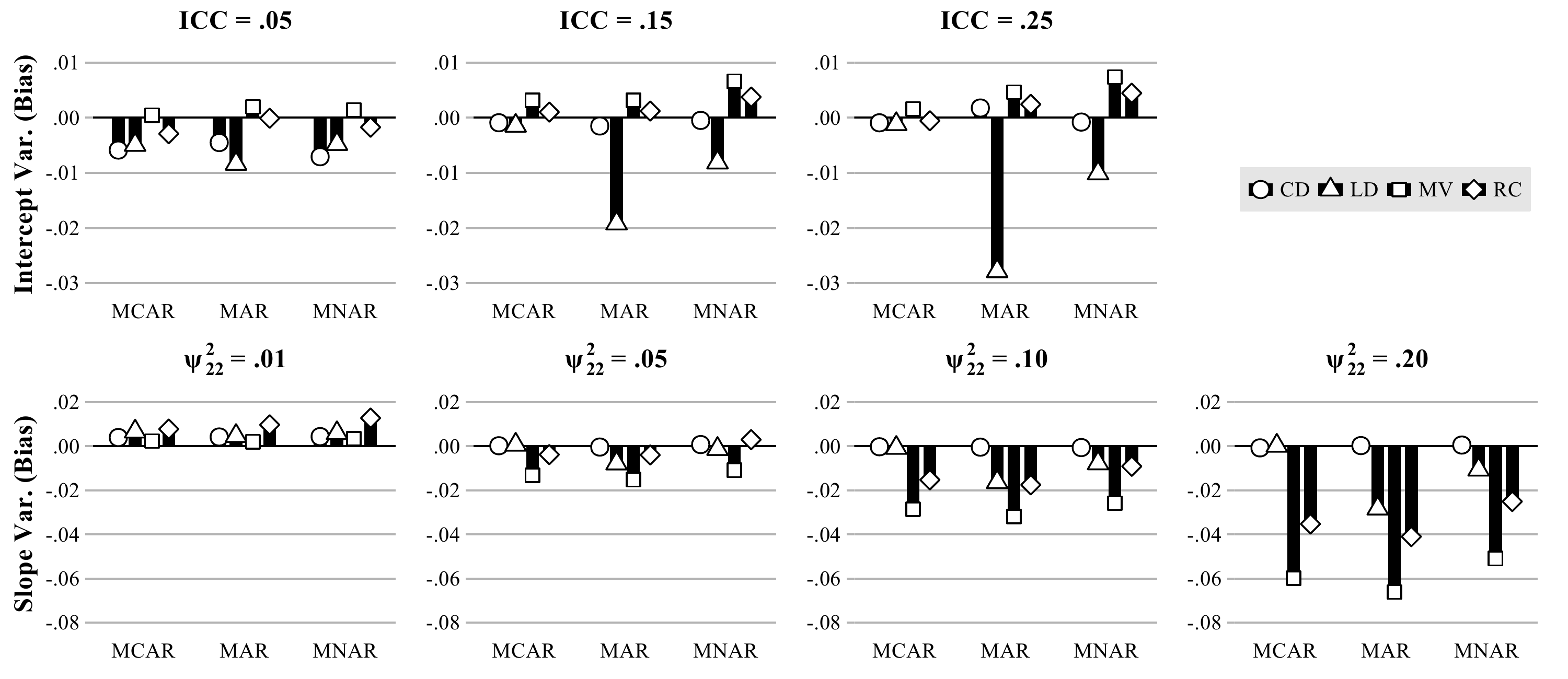}
  \caption{Bias in estimating the intercept (top row) and slope variance (bottom row) for univariate missing data on $X$ (Study 2) for different values of the ICC or slope variance, respectively, and different MD mechanisms and MD techniques. ICC = true intraclass correlation; $\psi_{22}^2$ = true slope variance; MCAR = missing completely at random; MAR = missing at random; MNAR = missing not at random; CD = complete data; LD = listwise deletion; MV = multivariate imputation; RC = conditional imputation using the reversed RC model.}
  \label{fig:varcomp}
\end{figure}

Conditional and multivariate MI underestimated the intercept variance when the ICCs were small but provided unbiased estimates otherwise.
Listwise deletion followed the same pattern for MCAR data, but otherwise underestimated the intercept variance.
This bias was strongest in the MAR condition, weaker with MNAR data, and increased as the ICCs grew larger.
Figure \ref{fig:varcomp} (top row) illustrates this finding for different levels of the ICC.

Results for the slope variance differed from Study 1.
Although conditional MI again overestimated small amounts of slope variation, this bias was much weaker and practically disappeared in larger samples (see Table \ref{tab:res2}).
Moderate slope variation could be estimated almost without bias.
In contrast to Study 1, however, large and very large slope variances were not estimated correctly by conditional MI but increasingly suffered from a downward bias.
Listwise deletion provided practically unbiased estimates of the slope variance if the sample size was sufficiently large.
The positive bias for conditional MI was also present with MNAR data, whereas the negative bias was smaller.
Figure \ref{fig:varcomp} (bottom row) illustrates these findings for different levels of slope variation.

According to the RMSE, the intercept variance could occasionally be estimated more efficiently using MI, whereas the slope variance could be estimated more accurately using LD.
However, these differences were usually very small.
Supplement D even suggests that conditional MI occasionally estimated the slope variance more efficiently in small samples.

\subsubsection{Other parameters}
The covariance between random intercepts and slopes was well recovered across all conditions.
The Level~1 residual variance was overestimated using MI, where conditional MI was less biased, but it was underestimated by LD when data were not MCAR.
For higher amounts of slope variation, the bias associated with LD became smaller, whereas the bias grew for MI.
These patterns were observed with MAR and MNAR data, but the bias was relatively small.
 
\subsection{Discussion} 
Regarding most parameters of the analyst's model, better estimates could be obtained using the reversed MI procedure, especially when the covariate $X$ was not MCAR.
This was true for the fixed regression coefficients but also applied to the intercept variance and even transferred to MNAR data.
However, reversed MI seemed to provide unstable estimates of the slope variance, which could be positively or negatively biased.
The positive bias for small slope variances became essentially zero as the samples grew larger.
For larger slope variances, the bias did not approach zero (as in Study 1) but turned negative regardless of sample size.
The negative bias was, however, rather small and could be viewed as negligible considering that it only occurred for large slope variances, which are rarely found in empirical studies.
Furthermore, the overall precision of the estimates, as indicated by the RMSE, was often comparable to LD because the data were handled more efficiently using MI.
The reversed model seemed to share many but not all of the desirable properties of the regular RC model.

The multivariate imputation model is applicable if little slope variation is present in the data, but it will suppress even moderate amounts of slope variation and inflate the Level~1 residual variance.
Estimates of the fixed slope obtained from multivariate MI were even less biased and more efficient than those from the reversed MI procedure.
Listwise deletion offered little benefit as most of its parameter estimates were biased unless the data were MCAR.
However, LD provided surprisingly accurate results for the slope variance.
Small variance components were again positively biased but less so than in the previous study.
We will return to this point in the General Discussion.

\section{Study 3}
The final study examined the performance of MI and LD with multivariate missing data.
The analyst's model was once again the RC model $f(Y|X,\boldsymbol\theta_X)$, but only the multivariate imputation model $g(X,Y|\boldsymbol\omega_0)$ could be applied.
This imputation model ignores slope variability, but may provide reasonable results for the remaining parameters of the analyst's model.

\subsection{Simulation and Methods}
The same procedures that were used in the previous studies could be used for most tasks.
Because the pattern of missing data was no longer univariate, the missing data model had to be adjusted.
We excluded unit-nonresponse from our considerations; thus, every participant was expected to have at least one observation on either $X$ or $Y$.
This allowed us to implement the same mechanisms as described before (i.e., MCAR, MAR, MNAR) for both $X$ and $Y$.
For each case, a coin toss decided whether $X$ or $Y$ could be missing (i.e., each was equally likely).
The actual missing values were then imposed on either $X$ or $Y$ with the probability that was given in the simulation design.
Thus the amount of missing values in each dataset was the same in all three studies.

\subsection{Results and Discussion}
\begin{table}[tbp]
  \begin{threeparttable}
    \caption{Study 3: Bias and RMSE for Estimates Obtained from LD and MI Given Small Variance Components, Smaller or Larger Samples, and Missing $X$ and $Y$ ~\\[1.8ex]}
    \label{tab:res3}
    \setlength{\tabcolsep}{5pt}
    \renewcommand{\arraystretch}{1.15}
    \small
    \begin{tabular}{lcccccccccccc}
\\[-5ex]\hline\\[-2.5ex]
 & \multicolumn{6}{c}{Bias} & \multicolumn{6}{c}{RMSE} \\ \cmidrule(lr){2-7}\cmidrule(lr){8-13}
 & \multicolumn{2}{c}{MCAR} & \multicolumn{2}{c}{MAR} & \multicolumn{2}{c}{MNAR} & \multicolumn{2}{c}{MCAR} & \multicolumn{2}{c}{MAR} & \multicolumn{2}{c}{MNAR} \\ \cmidrule(lr){2-3}\cmidrule(lr){4-5}\cmidrule(lr){6-7}\cmidrule(lr){8-9}\cmidrule(lr){10-11}\cmidrule(lr){12-13}
Est.  & LD & MV & LD & MV & LD & MV & LD & MV & LD & MV & LD & \multicolumn{1}{c}{MV} \\ 
[0ex]\hline\\[-2.8ex]
& \multicolumn{11}{c}{A: N=10, G=50, ICC=0.05, SV=0.01} \\[0ex] \hline \\[-2.7ex]$\beta_0^{}$  & $\phantom{-}.002$ & $\phantom{-}.001$ & $\mathbf{-.084}$ & $-.001$ & $\mathbf{-.097}$ & $-.019$ & $\phantom{-}.003$ & $\phantom{-}.003$ & $\mathbf{\phantom{-}.010}$ & $\phantom{-}.002$ & $\mathbf{\phantom{-}.012}$ & $\phantom{-}.003$ \\
$\beta_1^{}$  & $\phantom{-}.001$ & $-.001$ & $\mathbf{-.026}$ & $-.001$ & $\mathbf{-.048}$ & $-.024$ & $\phantom{-}.002$ & $\phantom{-}.002$ & $\phantom{-}.003$ & $\phantom{-}.002$ & $\mathbf{\phantom{-}.005}$ & $\phantom{-}.003$ \\
$\psi_{11}^2$  & $-.007$ & $\phantom{-}.005$ & $-.006$ & $\phantom{-}.006$ & $-.003$ & $\phantom{-}.006$ & $\phantom{-}.001$ & $\phantom{-}.001$ & $\phantom{-}.001$ & $\phantom{-}.001$ & $\phantom{-}.001$ & $\phantom{-}.001$ \\
$\psi_{22}^2$  & $\mathbf{\phantom{-}.006}$ & $\phantom{-}.002$ & $\mathbf{\phantom{-}.005}$ & $\phantom{-}.001$ & $\mathbf{\phantom{-}.005}$ & $\phantom{-}.001$ & $\phantom{-}.000$ & $\phantom{-}.000$ & $\phantom{-}.000$ & $\phantom{-}.000$ & $\phantom{-}.000$ & $\phantom{-}.000$ \\
$\psi_{12}^{}$  & $-.001$ & $-.002$ & $-.001$ & $-.001$ & $-.001$ & $-.001$ & $\phantom{-}.000$ & $\phantom{-}.000$ & $\phantom{-}.000$ & $\phantom{-}.000$ & $\phantom{-}.000$ & $\phantom{-}.000$ \\
$\sigma_{}^2$  & $-.001$ & $-.001$ & $-.026$ & $-.000$ & $-.025$ & $\phantom{-}.004$ & $\phantom{-}.003$ & $\phantom{-}.003$ & $\phantom{-}.004$ & $\phantom{-}.003$ & $\phantom{-}.004$ & $\phantom{-}.003$ \\
[0ex]\hline\\[-2.8ex] 
& \multicolumn{11}{c}{B: N=30, G=150, ICC=0.05, SV=0.01} \\[0ex] \hline \\[-2.7ex]$\beta_0^{}$  & $\phantom{-}.000$ & $\phantom{-}.000$ & $\mathbf{-.082}$ & $-.000$ & $\mathbf{-.099}$ & $-.020$ & $\phantom{-}.000$ & $\phantom{-}.000$ & $\mathbf{\phantom{-}.007}$ & $\phantom{-}.000$ & $\mathbf{\phantom{-}.010}$ & $\phantom{-}.001$ \\
$\beta_1^{}$  & $\phantom{-}.001$ & $\phantom{-}.000$ & $\mathbf{-.027}$ & $\phantom{-}.000$ & $\mathbf{-.046}$ & $-.020$ & $\phantom{-}.000$ & $\phantom{-}.000$ & $\mathbf{\phantom{-}.001}$ & $\phantom{-}.000$ & $\mathbf{\phantom{-}.002}$ & $\phantom{-}.001$ \\
$\psi_{11}^2$  & $-.002$ & $\phantom{-}.000$ & $-.003$ & $\phantom{-}.001$ & $-.003$ & $\phantom{-}.001$ & $\phantom{-}.000$ & $\phantom{-}.000$ & $\phantom{-}.000$ & $\phantom{-}.000$ & $\phantom{-}.000$ & $\phantom{-}.000$ \\
$\psi_{22}^2$  & $\phantom{-}.000$ & $\mathbf{-.003}$ & $-.001$ & $\mathbf{-.004}$ & $-.000$ & $\mathbf{-.004}$ & $\phantom{-}.000$ & $\phantom{-}.000$ & $\phantom{-}.000$ & $\phantom{-}.000$ & $\phantom{-}.000$ & $\phantom{-}.000$ \\
$\psi_{12}^{}$  & $-.000$ & $-.000$ & $-.001$ & $-.001$ & $-.001$ & $-.000$ & $\phantom{-}.000$ & $\phantom{-}.000$ & $\phantom{-}.000$ & $\phantom{-}.000$ & $\phantom{-}.000$ & $\phantom{-}.000$ \\
$\sigma_{}^2$  & $\phantom{-}.003$ & $\phantom{-}.005$ & $-.024$ & $\phantom{-}.004$ & $-.022$ & $\phantom{-}.008$ & $\phantom{-}.000$ & $\phantom{-}.000$ & $\mathbf{\phantom{-}.001}$ & $\phantom{-}.000$ & $\mathbf{\phantom{-}.001}$ & $\phantom{-}.000$ \\
[0ex]\hline\\[-2.8ex] 
\end{tabular}

    \begin{tablenotes}[para,flushleft] ~\\[-2.0ex]
      {\small \textit{Note.} LD = listwise deletion; MV = multivariate imputation; MCAR = missing completely at random; MAR = missing at random; MNAR = missing not at random; $\beta_0$ = fixed intercept; ; $\beta_1$ = fixed slope; $\psi_{11}^2$ = intercept variance; $\psi_{22}^2$ = slope variance; $\psi_{12}$ = intercept-slope covariance; $\sigma^2$ = Level~1 residual variance.} 
    \end{tablenotes}
  \end{threeparttable}
\end{table}
The results of the third study provided little further insight into the performance of LD and multivariate MI because the bias and RMSE were usually halfway between those reported in Studies 1 and 2.
Results for small variance components are presented in Table \ref{tab:res3}.
The complete results are available in Supplement D.
Multivariate MI provided approximately unbiased estimates of all parameters as long as the slope variance was close to zero and the values were either MCAR or MAR.
The slope variance was underestimated by as much as 40\%, especially in larger samples where more values were imputed under false assumptions.
When the data were MNAR, multivariate MI underestimated the fixed regression coefficient, but the bias was relatively small compared with the true values.
Estimates obtained from LD were approximately unbiased when data were MCAR.
When data were MAR or MNAR, the fixed effects were biased downward and were estimated less efficiently than with multivariate MI, where higher values for the ICC and slope variance reduced bias with LD (see Supplement D).

The results of the third study suggest that MI is necessary for proper estimation of the fixed regression coefficients.
Unfortunately, pan's multivariate imputation model could not preserve the slope variance.
If the slope variance was small and the number of missing values was not very high, then the bias was relatively small in absolute size.
Limiting the analysis to complete cases only distorted the parameter estimates, but provided reasonable estimates of the slope variance.

\section{General Discussion}
We investigated the performance of conditional and multivariate MI for univariate and multivariate patterns of missing data.
Both conditional MI and LD provided unbiased estimates if only the outcome was missing.
Care should be taken if covariates are partially unobserved.
Imputing the covariate in a reversed manner accounted for, but also misspecified the slope variation.
Only vague estimates could be obtained for the slope variance, but bias was not extreme, and the remaining estimates exhibited either no or less bias than what would have been obtained by deleting cases. 
The multivariate imputation model rarely induced any bias but strongly underestimated the slope variance.
Thus, it is appropriate only if the true slope variance is close to zero and not too many values are unobserved.
We recommend that LD be avoided when covariate data are missing unless the data are strictly MCAR.

As is true for all computer simulations, our study was limited in several ways.
The missing data mechanisms were based on linear models and may behave quite differently in nature.
Other implementations are possible, and results may vary especially for MAR and MNAR data \citep{Allison2000, Galati2013}.
We focused on descriptive measures of approximate performance but ignored statistical inference.
Testing for slope variation \citep{LaHuis2007} as well as Type-I and Type-II error rates associated with LD and MI should be a subject of future research.
Rather than estimating the slope variance, researchers often wish to explain it using predictor variables at Level~2 \citep{Mathieu2012, Aguinis2013}.
Cross-level interaction effects might be relatively easy to recover even if the slope variance is not.

Interestingly, small variance components were positively biased across the three studies.
We argue that this is due to the standard least-informative prior, which induces bias into small variance components.
Ad hoc procedures might combine the specific advantages of LD and MI and lead to less biased and more stable estimates.
For example, choosing $D^{-1} = 2 \cdot \hat{\boldsymbol\Psi}_{LD}$ as the scale matrix of the inverse Wishart prior for the covariance matrix of random effects, where $\hat{\boldsymbol\Psi}_{LD}$ is an estimate of this covariance matrix obtained from LD, would loosely center the prior distribution around appropriate values.
The computer code for this specification is provided in Supplement C of the supplemental online materials.
We conducted a small simulation to examine whether the bias for the intercept and slope variance could be reduced by rescaling the prior distribution in this manner.
The simulation featured small samples, univariate MAR data on either $X$ or $Y$, small values for the ICCs, as well as small and very large values for the slope variance.
Estimates of small variance components that utilized the adjusted prior did not exhibit any more bias than LD did and were often more efficient.
The positive bias reported in Studies 1 and 2 could therefore be viewed as an artifact of specifying the least-informative prior.
The negative bias for large slope variances in Study 2, however, could not be improved in this manner.
Using least squares or maximum likelihood estimation might further strengthen this approach.

The methodological literature offers alternatives to pan for multilevel MI.
It has been suggested that multilevel data be imputed using dummy variables in random intercept models but that imputations should be conducted separately for each group if random slopes are involved \citep{Graham2009, Graham2012}.
However, \citet{Andridge2011} found that the first approach leads to biased results, and unreported simulation results indicate that very large samples are needed to treat even small amounts of missing data with the second approach.
Alternative MI procedures include \emph{fully conditional specification} using chained equations \citep{vanBuuren2011}.
These procedures might lead to better results, but may face similar problems with respect to the slope variance.
However, recent developments in the context of substantive model compatible MI have offered promising results for interaction effects and nonlinear terms among covariates that have missing values \citep{vonHippel2009, Bartlett2014a}.
Extending this approach to multilevel MI \citep{Goldstein2009, Goldstein2014} and applying it to random slope models should be the subject of future research.
Adaptations of the pan model have been proposed by \citet{Shin2010} and \citet{Yucel2011}.
The latter approach specifies a joint model that allows the within-group covariance matrix to vary across groups, and has been recently discussed by \citet{Carpenter2013}.
However, it is currently not available in standard software and has yet to be evaluated in a systematic manner. 

In general, we believe that MI is a flexible and powerful tool that can be used to treat missing data in multilevel research.
More research should be conducted to generalize the current formulations of MI and to evaluate recent developments as well as sensible ad hoc solutions to missing data in multilevel models with random slopes. 

\bibliographystyle{apacite}
\bibliography{randomslopes_fullbib}

\end{document}